\documentclass[aps,prb,twocolumn,showpacs,groupedaddress]{revtex4-1}
\usepackage{amssymb}

\usepackage{graphicx}
\usepackage{subfig}

\begin{document}

\title[]{Weak Coupling BCS-like Superconductivity in the Pnictide Oxide Ba$_{1-x}$Na$_x$Ti$_2$Sb$_2$O (x=0 and 0.15)}

\author{Melissa Gooch$^1$, Phuong Doan$^2$, Zhongjia Tang$^2$, Bernd Lorenz$^1$, Arnold M. Guloy$^2$, and Paul C. W. Chu$^{1,3}$}

\affiliation{$^1$ TCSUH and Department of Physics, University of Houston, Houston, Texas 77204, USA}

\affiliation{$^2$ TCSUH and Department of Chemistry, University of Houston, Houston, Texas 77204, USA}

\affiliation{$^3$ Lawrence Berkeley National Laboratory, 1 Cyclotron Road, Berkeley, California 94720, USA}

\begin{abstract}
We report the results of low-temperature heat capacity measurements of the pnictide oxide superconductor BaTi$_2$Sb$_2$O doped with sodium. The temperature and field dependent heat capacity data are well described by a single-gap BCS theory. The estimated values for the normal state Sommerfeld constant, the heat capacity jump at $T_c$, and the electron-phonon coupling constant are in favor of a conventional weak coupling superconductivity mediated by electron-phonon interaction. The results are discussed with regard to and compared with recent first principle calculations.
\end{abstract}


\maketitle

\section{Introduction}
Superconductivity in layered transition metal compounds has been in the focus of interest for several decades. The discovery of high-temperature superconductivity in copper oxide perovskites as well as, more recently, in iron pnictides and chalcogenides has initiated an immense activity in superconductivity research. Questions about the role of the layered structure, magnetic fluctuations, the proximity to magnetic order or a possible quantum critical point, the symmetry of the superconducting order parameter, etc. have been discussed extensively. There are numerous examples showing that the superconducting state, usually induced by carrier doping into a parent compound, is in close proximity to spin or charge orders, for example antiferromagnetic order in copper oxide and heavy fermion superconductors, spin density wave (SDW) order in iron pnictides, or charge density wave (CDW) orders in layered chalcogenides.\cite{chu:10,paglione:10,norman:11,sipos:08,morosan:06} The mutual interactions of different fundamental orders open new perspectives to study novel physical phenomena only observable in those strongly correlated superconducting systems.

The search for similar layered transition metal compounds has recently led to the study of titanium-based pnictide oxide compounds of the general chemical formulas A$_2$Ti$_2$Pn$_2$O and AeTi$_2$Pn$_2$O (Pn = As, Sb; A = Na, Ae = Ba). This system was first synthesized and characterized more than two decades ago\cite{adam:90} and it belongs to a larger family of pnictide oxides including other transition metals.\cite{brock:95,ozawa:08,johrendt:11} The layered structure of Na$_2$Ti$_2$Pn$_2$O is characterized by Ti$_2$Pn$_2$O blocks separated by a double layer of A (Na) ions or single slabs of Ae (Ba). The titanium and oxygen ions form square planar Ti$_2$O sheets which is inverse (anti-) to the CuO$_2$ planes which form the active layer in superconducting copper oxides, in that the positions of Ti and O in this square lattice are switched with respect to the O and Cu sites in the cuprates. This raised the question whether or not a superconducting state can be induced through doping of carriers into the Ti$_2$O layer.\cite{ozawa:08}

The structure and chemical synthesis of Na$_2$Ti$_2$Pn$_2$O have been studied in detail for Pn=As and Sb.\cite{adam:90,ozawa:00,ozawa:04} The physical properties studied via electrical transport and magnetic measurements reveal the existence of an instability near $T_{DW}$=115 K in Na$_2$Ti$_2$Sb$_2$O where the magnetic susceptibility shows a sharp drop to lower values and the resistivity experiences a sudden increase.\cite{axtell:97,ozawa:04,liu:09} However, it is not clear if this instability originates from the formation of a charge density wave or a spin density wave. First principle calculations have found a nesting feature of the Fermi surface which could explain the occurrence of a SDW or CDW instability.\cite{pickett:98} Powder neutron scattering experiments have shown that this transition is also associated with a small structural anomaly.\cite{ozawa:00} Recent optical studies have found the opening of a gap and a change of the free carrier spectral weight as well as a reduction of the carrier scattering rate across the SDW/CDW transition.\cite{huang:13} It is worth noting that the arsenic-based compound, Na$_2$Ti$_2$As$_2$O exhibits a similar density wave transition at significantly higher temperature, $T_{DW}$=320 K.\cite{liu:09}

Replacing the Na$_2$ double layer with a single layer of Ba ions forms another stable pnictide oxide, BaTi$_2$As$_2$O, with the c-axis of the tetragonal structure shortened by 50 \%.\cite{wang:10} Interestingly, this compound features identical Ti$_2$As$_2$O layers and it also exhibits the density wave transition at $T_{DW}$=200 K, albeit significantly lower than in the related compound Na$_2$Ti$_2$As$_2$O. However, attempts to induce superconductivity in BaTi$_2$As$_2$O through doping or intercalation have not been successful.\cite{wang:10}

Replacing As by Sb and forming BaTi$_2$Sb$_2$O was only achieved very recently.\cite{doan:12,yajima:12} Superconductivity was observed below 1.2 K\cite{yajima:12} and the critical temperature could be raised to above 5 K through Na doping at the Ba site in Ba$_{1-x}$Na$_x$Ti$_2$Sb$_2$O.\cite{doan:12} The undoped parent compound shows a density wave transition at much lower temperature, $T_{DW}$=54 K, and it was shown that Na-doping further decreased the critical temperature even further (while raising the superconducting $T_c$).\cite{doan:12} The question about the nature of the superconducting state and whether or not the density wave state competes or supports superconductivity is yet to be explored.

Electronic structure calculations by Singh\cite{singh:12} have shown the existence of several sheets at the Fermi surface and a nesting property which induces a magnetic instability (SDW). Related to this instability, a sign-changing s-wave state was predicted to originate from an unconventional spin fluctuation mediated superconductivity. The coupling of charges to the magnetic order is observed in a strong magnetoresistance effect below $T_{DW}$ in BaTi$_2$Sb$_2$O\cite{doan:12} as well as in BaTi$_2$As$_2$O.\cite{wang:10} On the other hand, Subedi\cite{subedi:13} suggested the electron-phonon mechanism for superconductivity and a CDW instability in BaTi$_2$Sb$_2$O based on first-principles calculations of the phonon dispersions and electron-phonon interactions, predicting a multiband superconducting state, similar to some iron pnictide superconductors.

To distinguish between different proposals about the microscopic origin of superconductivity in Ba$_{1-x}$Na$_x$Ti$_2$Sb$_2$O experimental probes on the superconducting gap structure need to be employed. The temperature dependence of the electronic heat capacity, C$_{el}$(T), measured to low temperatures is a sensitive probe which can reveal unconventional superconductivity, the existence of nodes in the superconducting gap function, and the possible presence of multiple gaps (multiband superconductivity). We have therefore studied the heat capacity of Ba$_{1-x}$Na$_x$Ti$_2$Sb$_2$O in applied magnetic fields and at temperatures as low as 0.4 K. Our results are consistent with a single-gap weak coupling BCS-like superconducting state.

\section{Experimental}
High purity polycrystalline samples of BaTi$_2$Sb$_2$O was synthesized by reacting stoichiometric amounts of BaO (99.99$\%$ Aldrich) and Ti (99.99$\%$ Aldrich), Sb pieces (99.999$\%$, Alfa Aesar ) in welded niobium containers within evacuated quartz jackets.  The well mixed reactants were pressed into pellets and then sealed in niobium tubes under argon and subsequently jacketed, evacuated and sealed in quartz tubes. The samples
were heated slowly (2¼C/min) to 900¼C and annealed for 3 days, then slowly cooled (1 ¼C/min) to
200¼C.  To ensure phase homogeneity, an additional regrinding and sintering at 900¼C for another 3 days was performed.  Ba$_{0.85}$Na$_{0.15}$Ti$_2$Sb$_2$O was synthesized by reacting stoichiometric amounts of BaO (99.99$\%$, Sigma Aldrich); BaO$_2$ (95$\%$ , Sigma Aldrich); Na$_2$O (80$\%$ Na$_2$O and 20$\%$ Na$_2$O$_2$, Sigma Aldrich); Ti (99.99$\%$, Sigma Aldrich), Sb (pieces, 99.999$\%$, Alfa Aesar) in welded niobium containers within evacuated quartz jackets. The major impurity in Na$_2$O (20$\%$ Na$_2$O$_2$) had to be taken into account during the weighing.  The well mixed reactants were pressed into pellets and sealed in niobium tubes under argon, and subsequently jacketed, evacuated and sealed in quartz
tubes. The samples was heated slowly (2¼C/min) to 900¼C and annealed for 3 days, then slow cooled (1
¼C/min) to 200¼C.  Same as the BaTi$_2$Sb$_2$O, an additional regrinding and sintering at 900¼C for another 3 days was performed to ensure phase homogeneity.  The phase purity of the samples was investigated through powder X-ray diffraction, as well through, inductively coupled plasma/mass spectrometry (ICP-MS).  ICP-MS confirmed the Ba:Na molar ratio to be close to the nominal composition, as reported earlier.\cite{doan:12}  To prevent contamination of the air and moisture sensitive material all preparations had to be performed in an argon glove box.  The magnetic properties were measured using a Magnetic Property Measurement System (Quantum Design) and the heat capacity data was collected using a relaxation method and the He3-probe attached to the Physical Property Measurement System (Quantum Design).  Four probe resistivity measurements were carried out in the Physical Property Measurement System He-3 probe using an external LR700 (Linear Research) low-frequency (19 Hz) ac bridge. Measurements were carried out to temperatures as low as 0.4 K and in magnetic fields up to 70 kOe.

\begin{figure}
\begin{center}
\subfloat a){\label{a}\includegraphics[angle=0, width=2.5 in]{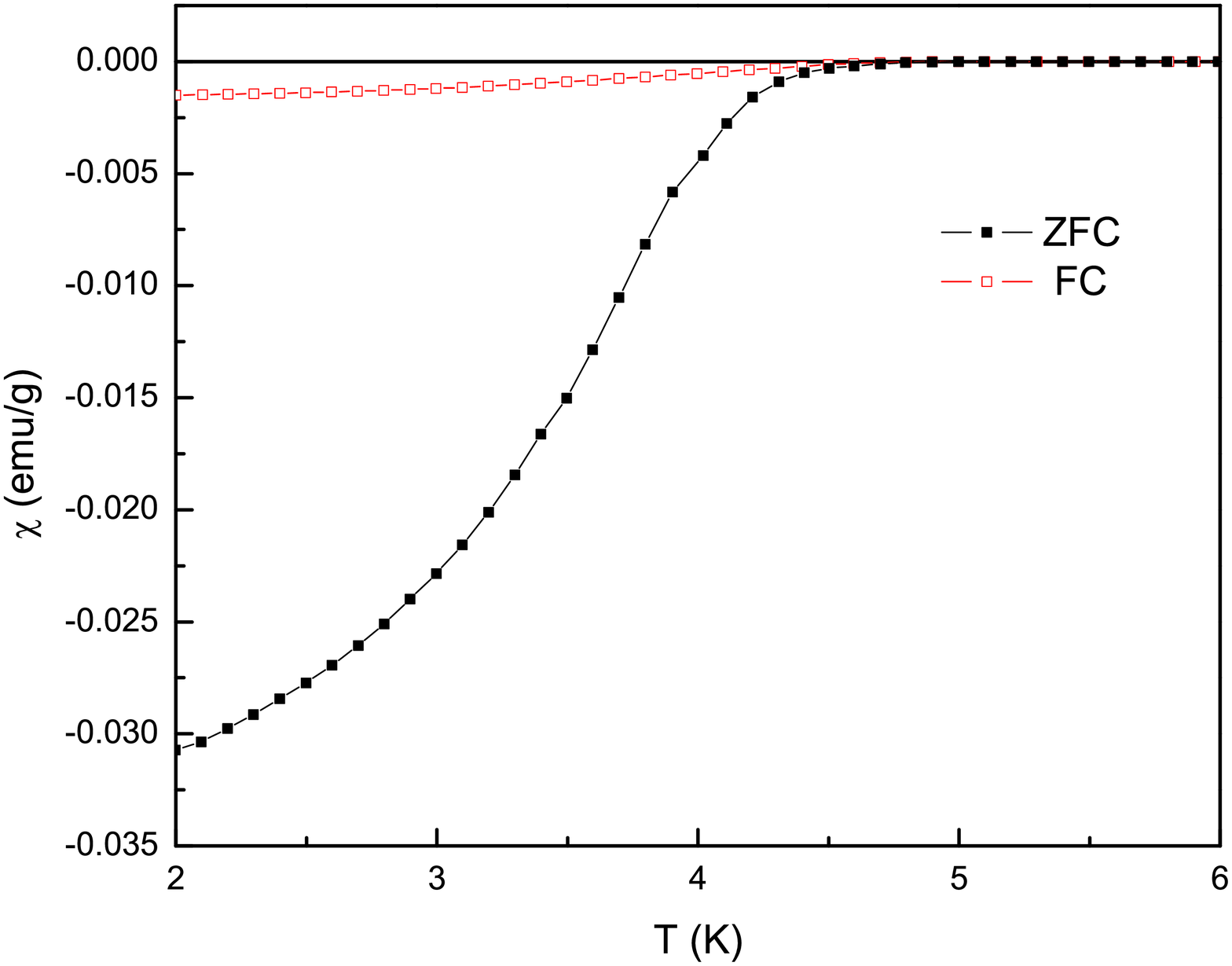}}
\subfloat b){\label{b}\includegraphics[angle=0, width=2.5 in]{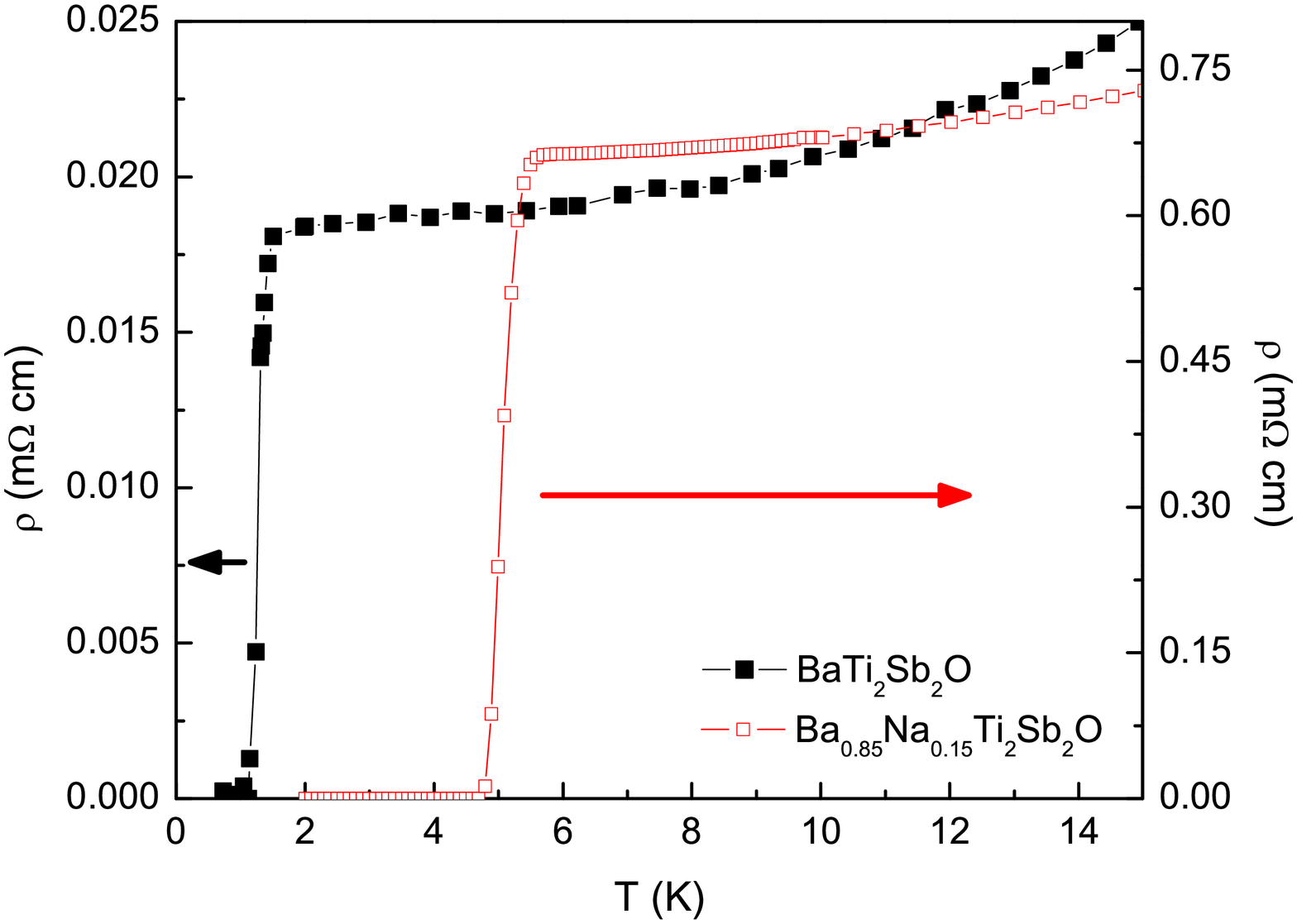}}
\end{center}
\caption{(Color online) a) Magnetic susceptibility $\chi(T)$ of Ba$_{0.85}$Na$_{0.15}$Ti$_2$Sb$_2$O measured in field cooling (FC) and zero-field cooling (ZFC) conditions. b) Resistivity of BaTi$_2$Sb$_2$O and Ba$_{0.85}$Na$_{0.15}$Ti$_2$Sb$_2$O}
\end{figure}

\begin{figure}
\begin{center}
\includegraphics[angle=0, width=2.5 in]{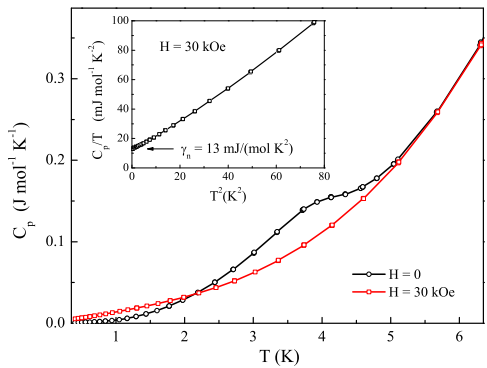}
\end{center}
\caption{(Color online) Heat capacity of Ba$_{0.85}$Na$_{0.15}$Ti$_2$Sb$_2$O measured in zero field and at 30 kOe (above $H_{c2}$). The inset shows the normal state heat capacity $C_p$/T vs. $T^2$ (30 kOe) data). The Sommerfeld constant is determined as $\gamma_n$=13 mJ/(mol K$^2$).}
\end{figure}
\section{Results and Discussions}
The superconducting transition for BaTi$_2$Sb$_2$O and Ba$_{0.85}$Na$_{0.15}$Ti$_2$Sb$_2$O is clearly indicated in Fig. 1a and Fig. 1b.  Ba$_{0.85}$Na$_{0.15}$Ti$_2$Sb$_2$O superconducting transition is seen in Fig. 1a, by the diamagnetic signal with onset below 5 K.  This transition is also confirmed through resistivity data, as seen in Fig. 1b, with the midpoint of the transition occurring at 5 K.  Fig 1b also shows resistivity data for BaTi$_2$Sb$_2$O. The superconducting transition is very sharp and the midpoint of the transition is at 1.2 K. The sharpness of the transition is an indication of the high purity of the sample which is further revealed by a high residual resistance ratio of about 40, significantly higher than in previous reports.\cite{yajima:12,zhai:13} 

The heat capacity data of Ba$_{0.85}$Na$_{0.15}$Ti$_2$Sb$_2$O are shown in Fig. 2. The superconducting state is completely suppressed in a magnetic field of 30 kOe. From the high-field data the normal state Sommerfeld constant is obtained as $\gamma_n$=13 mJ/(mol K$^2$). However, it should be noted that this value for $\gamma_n$ is enhanced by the electron-phonon coupling constant, as discussed below, and the bare electronic value relating to the density of states is lower.

The electronic heat capacity, $C_{el}(T)$ was derived from the data shown in Fig. 2 by subtracting the lattice contribution determined from the high-field data. The resulting data for the normalized $C_{el}/(\gamma_n T)$ are very well described by weak-coupling BCS function, shown as the dashed line in Fig. 3. The excellent agreement of the data with the BCS theory is evidence for conventional superconductivity in the pnictide oxide family of compounds. In addition, we do not find any signature of a possible multiband superconducting state which would leave the typical imprint on the low-temperature part of the heat capacity as, for example, observed in MgB$_2$.\cite{bouquet:01,lorenz:06} The superconducting gap value is estimated as $2\Delta$ = 1.3 meV.

\begin{figure}
\begin{center}
\includegraphics[angle=0, width=2.5 in]{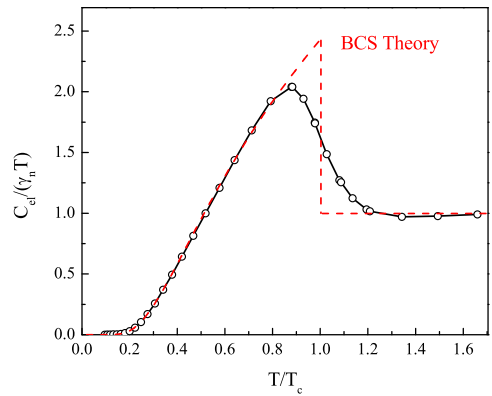}
\end{center}
\caption{(Color online) Normalized electronic heat capacity $C_{el}/(\gamma_n T)$ of Ba$_{0.85}$Na$_{0.15}$Ti$_2$Sb$_2$O measured in zero field. The dashed line represents the heat capacity from weak-coupling BCS theory.}
\end{figure}

The magnetic field dependence of $C_{el}(T)$ is shown in Fig. 4. The upper critical field $H_{c2}(T)$ is extracted from the shift of $T_c$ in magnetic fields, with $T_c$ defined as the midpoint of the sharp increase of $C_{el}(T)$. $H_{c2}(T)$ is shown in the inset of Fig. 4 and compared with the Werthamer-Helfand-Hohenberg (WHH) theory\cite{werthamer:66} (line). The zero temperature limit of the critical field is extrapolated as $H_{c2}(0)$ = 17 kOe, which results in a Ginzburg-Landau coherence length of $\xi(0)$ = 14 nm.

\begin{figure}
\begin{center}
\includegraphics[angle=0, width=2.5 in]{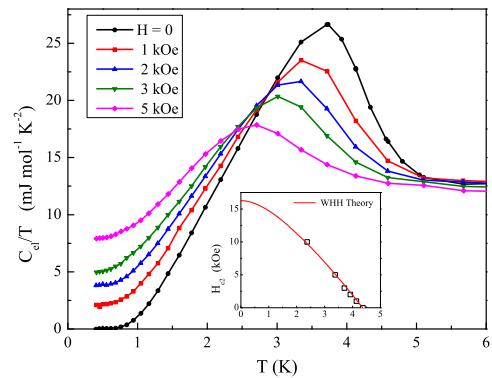}
\end{center}
\caption{(Color online) Magnetic field dependence of the electronic heat capacity $C_{el}/T$ of Ba$_{0.85}$Na$_{0.15}$Ti$_2$Sb$_2$O. The inset shows the upper critical field $H_{c2}(T)$ and the curve expected from the WHH theory.}
\end{figure}

The superconducting transition in the undoped BaTi$_2$Sb$_2$O happens at significantly lower temperature. However, a very broad transition reported originally\cite{yajima:12} provides a challenge for a detailed analysis of the low-temperature heat capacity. Our samples of BaTi$_2$Sb$_2$O were prepared with special care to reduce any impurity or defect content to a minimum.\cite{doan:12} As a result, a very sharp resistive transition at $T_c$=1.2 K (transition width $<$ 0.2 K) and also at $T_{DW}$ was observed as well as a much sharper heat capacity change at $T_c$. The zero-field normalized electronic heat capacity $C_{el}/(\gamma_n T)$ of BaTi$_2$Sb$_2$O is shown in Fig. 5 as function of the normalized temperature $T/T_c$. Unlike in the Na-doped compound discussed above, the BCS theory (dashed line in Fig. 5) does not fit the data well. Varying the superconducting gap size while maintaining the BCS-like temperature dependence of the gap (so called $\alpha$-model\cite{padamsee:73}) a much better fit is obtained with $\alpha\approx$ 1.4 (solid line in Fig. 5). These data still support a single-gap s-wave superconductivity with a gap value of $2\Delta$=0.3 meV, consistent with conclusions from recent NMR measurements.\cite{kitagawa:13} Interestingly, the heat capacity jump at $T_c$ is only 0.9 of the normal state value, much smaller than 1.43 (BCS theory) or 1.36 (Yajima et al.).\cite{yajima:12} This low value may suggest strong correlations, possibly related to the ordered density wave phase below $T_{DW}$. The Sommerfeld coefficient $\gamma_n$=10.9 mJ/(mol K$^2$) is about 20\% lower than reported earlier\cite{yajima:12} which we attribute to differences in sample purity.

The heat capacity peak in magnetic fields shows the expected shift to lower temperatures; however, the upper critical field is very low. Fig. 6 shows heat capacity data of BaTi$_2$Sb$_2$O in external fields up to 500 Oe. The upper critical field $H_{C2}(T)$ is shown in the inset of Fig. 6. The extrapolation to zero temperature, using the WHH theory, results in a very low value of $H_{c2}(0)\approx$ 800 Oe and a coherence length of $\xi(0)$=64 nm.

\begin{figure}
\begin{center}
\includegraphics[angle=0, width=2.5 in]{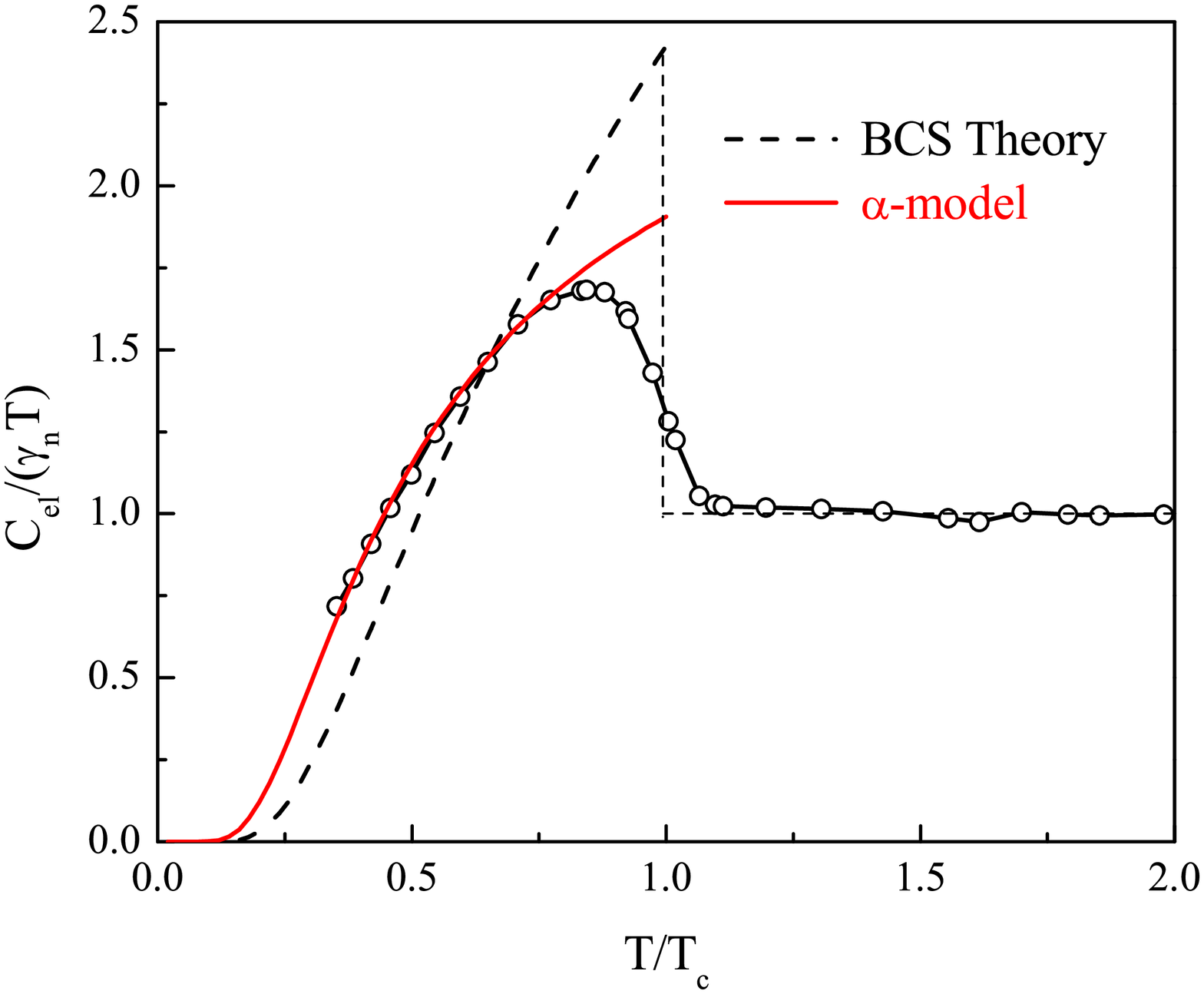}
\end{center}
\caption{(Color online) Normalized electronic heat capacity of BaTi$_2$Sb$_2$O at zero magnetic field. The BCS data are shown by the dashed line. The experimental data fit well to the $\alpha$-model (with $\alpha \approx$ 1.4).}
\end{figure}

\begin{figure}
\begin{center}
\includegraphics[angle=0, width=2.5 in]{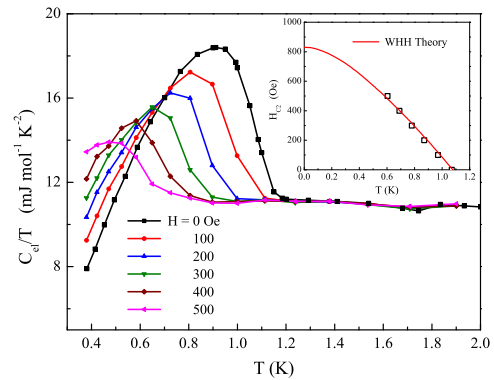}
\end{center}
\caption{(Color online) Magnetic field dependence of the electronic heat capacity $C_{el}/T$ of BaTi$_2$Sb$_2$O. The inset shows the upper critical field $H_{c2}(T)$ and the curve expected from the WHH theory.}
\end{figure}

Assuming the origin of the single-gap superconducting state lies in the electron-phonon coupling mechanism, as suggested by Subedi,\cite{subedi:13} the coupling constant $\lambda_{ep}$ and the bare electronic Sommerfeld constant $\gamma_0$ can be estimated from the McMillan formula for $\lambda_{ep}$\cite{mcmillan:68} and the relation $\gamma_n$=$\gamma_0$(1+$\lambda_{ep}$). With an assumed value of $\mu^*$=0.13, the results for x=0 and x=0.15 are summarized in Table 1. The Na doping results in a moderate increase of the density of states (as derived from the increase of $\gamma_0$ by about 5 \%), however, the electron-phonon coupling constant increases significantly by almost 40 \%, resulting in the four times higher value of $T_c$. However, it should be noted that this discussion is based solely on the assumption that the pairing mechanism is via the electron-phonon interaction,\cite{subedi:13} but alternative possibilities, e.g. spin fluctuation mediated pairing suggested by Singh,\cite{singh:12} cannot be ruled out.


\begin{table}[]
\caption{Superconducting and normal state parameters of Ba$_{1-x}$Na$_x$Ti$_2$Sb$_2$O. The units of $\gamma_n$ and $\gamma_0$ are mJ/(mol K$^2$).}
\vspace{0.2cm}
\centering
\begin{tabular}{@{\hspace{0.25cm}} c @{\hspace{0.25cm}}|@{\hspace{0.25cm}} c @{\hspace{0.25cm}}|@{\hspace{0.25cm}} c @{\hspace{0.25cm}}|@{\hspace{0.25cm}} c @{\hspace{0.25cm}}|@{\hspace{0.25cm}} c @{\hspace{0.25cm}}|@{\hspace{0.25cm}} c @{\hspace{0.25cm}}|@{\hspace{0.25cm}} c @{\hspace{0.25cm}}|@{\hspace{0.25cm}} r@{\hspace{0.25cm}}}
\hline
x&$T_c$&$H_{c2}$&$\xi$(0)&$\Theta_D$&$\lambda_{ep}$&$\gamma_n$&$\gamma_0$\\
 & K &kOe& \AA & K &&\\
\hline
 0 &1.1&0.8&640&230&0.49&10.9&7.3\\
0.15&4.2&17&140&210&0.68&13.0&7.7\\
\hline
\end{tabular}
\end{table}

\section{Conclusion}
The heat capacity data of BaTi$_2$Sb$_2$O and Ba$_{1-x}$Na$_x$Ti$_2$Sb$_2$O are best described by a single gap BCS function in the weak coupling limit. This provides evidence that the superconductivity in this compound is most likely of s-wave character, as was also implied from NMR/NQR measurements of the undoped compound.\cite{kitagawa:13} The question of whether the pairing is mediated by phonons or spin fluctuations is yet unresolved. The key seems to be the understanding of the nature of the density wave order below $T_{DW}$.

All theoretical studies so far agree about the nesting feature of the Fermi surface of Na$_2$Ti$_2$Sb$_2$O\cite{pickett:98,biani:98} and Ba$_{1-x}$Na$_x$Ti$_2$Sb$_2$O\cite{singh:12} which is most likely the origin of the density wave order. This order coexists with superconductivity in Ba$_{1-x}$Na$_x$Ti$_2$Sb$_2$O (0$\leq$x$\leq$0.33).\cite{yajima:12,doan:12,kitagawa:13} While this density wave order has been observed earlier in different pnictide oxides like Na$_2$Ti$_2$As$_2$O ($T_{DW}$=320 K)\cite{liu:09}, Na$_2$Ti$_2$Sb$_2$O ($T_{DW}$=114 K)\cite{axtell:97}, and BaTi$_2$As$_2$O ($T_{DW}$=200 K)\cite{wang:10}, no superconductivity was found in these systems and attempts of doping or intercalation have not been successful.\cite{wang:10} The superconductivity in Ba$_{1-x}$Na$_x$Ti$_2$Sb$_2$O appears in the compound with the lowest $T_{DW}\leq$54 K suggesting that density wave and superconducting states are competing with one another. The decrease of $T_{DW}$ and simultaneous increase of the superconducting $T_c$ upon Na doping\cite{doan:12} support the latter conclusion. Furthermore, the recently reported superconductivity in the system BaTi$_2$(Sb$_{1-x}$Bi$_x$)$_2$O also shows that the substitution of Bi for Sb results in a suppression of the density wave phase and an increase of $T_c$.\cite{yajima:13,zhai:13} Therefore, it is most likely that the superconducting state and the density wave order compete and the superconductivity is only possible in pnictide oxide compounds with a lower energy scale of the DW phase.  While our results suggest that the superconducting state is a conventional s-wave, the nature of the pairing mechanisms is unclear at this time and may be either induced by the electron-phonon coupling\cite{subedi:13} or spin fluctuation derived\cite{singh:12}. While no conclusion can be drawn on the pairing mechanism, we also have not found any evidence for a multiband character of the superconducting state in our heat capacity data.

\begin{acknowledgments}
This work is supported in part by the US Air Force Office of Scientific Research, the Robert A. Welch Foundation (E-1297), the T.L.L. Temple Foundation, the J. J. and R. Moores Endowment, and the State of Texas through the TCSUH and at LBNL by the DoE.
\end{acknowledgments}


\begin{thebibliography}{10}%
\makeatletter
\providecommand \@ifxundefined [1]{%
 \ifx #1\undefined \expandafter \@firstoftwo
 \else \expandafter \@secondoftwo
\fi
}%
\providecommand \@ifnum [1]{%
 \ifnum #1\expandafter \@firstoftwo
 \else \expandafter \@secondoftwo
\fi
}%
\providecommand \enquote [1]{``#1''}%
\providecommand \bibnamefont  [1]{#1}%
\providecommand \bibfnamefont [1]{#1}%
\providecommand \citenamefont [1]{#1}%
\providecommand\href[0]{\@sanitize\@href}%
\providecommand\@href[1]{\endgroup\@@startlink{#1}\endgroup\@@href}%
\providecommand\@@href[1]{#1\@@endlink}%
\providecommand \@sanitize [0]{\begingroup\catcode`\&12\catcode`\#12\relax}%
\@ifxundefined \pdfoutput {\@firstoftwo}{%
 \@ifnum{\z@=\pdfoutput}{\@firstoftwo}{\@secondoftwo}%
}{%
 \providecommand\@@startlink[1]{\leavevmode}%
 \providecommand\@@endlink[0]{}%
}{%
 \providecommand\@@startlink[1]{%
  \leavevmode
  \pdfstartlink
   attr{/Border[0 0 1 ]/H/I/C[0 1 1]}%
   user{/Subtype/Link/A<</Type/Action/S/URI/URI(#1)>>}%
  \relax
 }%
 \providecommand\@@endlink[0]{\pdfendlink}%
}%
\providecommand \url  [0]{\begingroup\@sanitize \@url }%
\providecommand \@url [1]{\endgroup\@href {#1}{\urlprefix}}%
\providecommand \urlprefix [0]{URL }%
\providecommand \Eprint[0]{\href }%
\@ifxundefined \urlstyle {%
  \providecommand \doi [1]{doi:\discretionary{}{}{}#1}%
}{%
  \providecommand \doi [0]{doi:\discretionary{}{}{}\begingroup
  \urlstyle{rm}\Url }%
}%
\providecommand \doibase [0]{http://dx.doi.org/}%
\providecommand \Doi[1]{\href{\doibase#1}}%
\providecommand \bibAnnote [3]{%
  \BibitemShut{#1}%
  \begin{quotation}\noindent
    \textsc{Key:}\ #2\\\textsc{Annotation:}\ #3%
  \end{quotation}%
}%
\providecommand \bibAnnoteFile [2]{%
  \IfFileExists{#2}{\bibAnnote {#1} {#2} {\input{#2}}}{}%
}%
\providecommand \typeout [0]{\immediate \write \m@ne }%
\providecommand \selectlanguage [0]{\@gobble}%
\providecommand \bibinfo [0]{\@secondoftwo}%
\providecommand \bibfield [0]{\@secondoftwo}%
\providecommand \translation [1]{[#1]}%
\providecommand \BibitemOpen[0]{}%
\providecommand \bibitemStop [0]{}%
\providecommand \bibitemNoStop [0]{.\EOS\space}%
\providecommand \EOS [0]{\spacefactor3000\relax}%
\providecommand \BibitemShut [1]{\csname bibitem#1\endcsname}%
\bibitem{chu:10}%
  \BibitemOpen
  \bibfield{author}{%
  \bibinfo {author} {\bibfnamefont{C.~W.}\ \bibnamefont{Chu}},\ }%
  \bibfield{journal}{%
  \bibinfo {journal} {Int. J. Mod. Phys. B}\ }%
  \textbf{\bibinfo {volume} {24}},\ \bibinfo {pages} {4102} (\bibinfo {year}
  {2010})%
  \bibAnnoteFile{NoStop}{chu:10}%
\bibitem{paglione:10}%
  \BibitemOpen
  \bibfield{author}{%
  \bibinfo {author} {\bibfnamefont{J.}~\bibnamefont{Paglione}}\ and\ \bibinfo
  {author} {\bibfnamefont{R.~L.}\ \bibnamefont{Greene}},\ }%
  \bibfield{journal}{%
  \bibinfo {journal} {Nature Physics}\ }%
  \textbf{\bibinfo {volume} {6}},\ \bibinfo {pages} {645} (\bibinfo {year}
  {2010})%
  \bibAnnoteFile{NoStop}{paglione:10}%
\bibitem{norman:11}%
  \BibitemOpen
  \bibfield{author}{%
  \bibinfo {author} {\bibfnamefont{M.~R.}\ \bibnamefont{Norman}},\ }%
  \bibfield{journal}{%
  \bibinfo {journal} {Science}\ }%
  \textbf{\bibinfo {volume} {332}},\ \bibinfo {pages} {196} (\bibinfo {year}
  {2011})%
  \bibAnnoteFile{NoStop}{norman:11}%
\bibitem{sipos:08}%
  \BibitemOpen
  \bibfield{author}{%
  \bibinfo {author} {\bibfnamefont{B.}~\bibnamefont{Sipos}}, \bibinfo {author}
  {\bibfnamefont{A.~F.}\ \bibnamefont{Kusmartseva}}, \bibinfo {author}
  {\bibfnamefont{A.}~\bibnamefont{Akrap}}, \bibinfo {author}
  {\bibfnamefont{H.}~\bibnamefont{Berger}}, \bibinfo {author}
  {\bibfnamefont{L.}~\bibnamefont{Forro}},\ and\ \bibinfo {author}
  {\bibfnamefont{E.}~\bibnamefont{Tutis}},\ }%
  \bibfield{journal}{%
  \bibinfo {journal} {Nature Materials}\ }%
  \textbf{\bibinfo {volume} {7}},\ \bibinfo {pages} {950} (\bibinfo {year}
  {2008})%
  \bibAnnoteFile{NoStop}{sipos:08}%
\bibitem{morosan:06}%
  \BibitemOpen
  \bibfield{author}{%
  \bibinfo {author} {\bibfnamefont{E.}~\bibnamefont{Morosan}}, \bibinfo
  {author} {\bibfnamefont{H.~W.}\ \bibnamefont{Zandbergen}}, \bibinfo {author}
  {\bibfnamefont{B.~S.}\ \bibnamefont{Dennis}}, \bibinfo {author}
  {\bibfnamefont{J.~W.~G.}\ \bibnamefont{Bos}}, \bibinfo {author}
  {\bibfnamefont{Y.}~\bibnamefont{Onose}}, \bibinfo {author}
  {\bibfnamefont{T.}~\bibnamefont{Klimczuk}}, \bibinfo {author}
  {\bibfnamefont{A.~P.}\ \bibnamefont{Ramirez}}, \bibinfo {author}
  {\bibfnamefont{N.~P.}\ \bibnamefont{Ong}},\ and\ \bibinfo {author}
  {\bibfnamefont{R.~J.}\ \bibnamefont{Cava}},\ }%
  \bibfield{journal}{%
  \bibinfo {journal} {Nature Physics}\ }%
  \textbf{\bibinfo {volume} {2}},\ \bibinfo {pages} {544} (\bibinfo {year}
  {2006})%
  \bibAnnoteFile{NoStop}{morosan:06}%
\bibitem{adam:90}%
  \BibitemOpen
  \bibfield{author}{%
  \bibinfo {author} {\bibfnamefont{A.}~\bibnamefont{Adam}}\ and\ \bibinfo
  {author} {\bibfnamefont{H.-U.}\ \bibnamefont{Schuster}},\ }%
  \bibfield{journal}{%
  \bibinfo {journal} {Z. Anorg. Allg. Chem.}\ }%
  \textbf{\bibinfo {volume} {584}},\ \bibinfo {pages} {150} (\bibinfo {year}
  {1990})%
  \bibAnnoteFile{NoStop}{adam:90}%
\bibitem{brock:95}%
  \BibitemOpen
  \bibfield{author}{%
  \bibinfo {author} {\bibfnamefont{S.~L.}\ \bibnamefont{Brock}}\ and\ \bibinfo
  {author} {\bibfnamefont{S.~M.}\ \bibnamefont{Kauzlarich}},\ }%
  \bibfield{journal}{%
  \bibinfo {journal} {Comments Inorg. Chem.}\ }%
  \textbf{\bibinfo {volume} {17}},\ \bibinfo {pages} {213} (\bibinfo {year}
  {1995})%
  \bibAnnoteFile{NoStop}{brock:95}%
\bibitem{ozawa:08}%
  \BibitemOpen
  \bibfield{author}{%
  \bibinfo {author} {\bibfnamefont{T.~C.}\ \bibnamefont{Ozawa}}\ and\ \bibinfo
  {author} {\bibfnamefont{S.~M.}\ \bibnamefont{Kauzlarich}},\ }%
  \bibfield{journal}{%
  \bibinfo {journal} {Sci. Technol. Adv. Mater.}\ }%
  \textbf{\bibinfo {volume} {9}},\ \bibinfo {pages} {033003} (\bibinfo {year}
  {2008})%
  \bibAnnoteFile{NoStop}{ozawa:08}%
\bibitem{johrendt:11}%
  \BibitemOpen
  \bibfield{author}{%
  \bibinfo {author} {\bibfnamefont{D.}~\bibnamefont{Johrendt}}, \bibinfo
  {author} {\bibfnamefont{H.}~\bibnamefont{Hosono}}, \bibinfo {author}
  {\bibfnamefont{R.-D.}\ \bibnamefont{Hoffmann}},\ and\ \bibinfo {author}
  {\bibfnamefont{R.}~\bibnamefont{Pottgen}},\ }%
  \bibfield{journal}{%
  \bibinfo {journal} {Z, Kristallogr.}\ }%
  \textbf{\bibinfo {volume} {226}},\ \bibinfo {pages} {435} (\bibinfo {year}
  {2011})%
  \bibAnnoteFile{NoStop}{johrendt:11}%
\bibitem{ozawa:00}%
  \BibitemOpen
  \bibfield{author}{%
  \bibinfo {author} {\bibfnamefont{T.~C.}\ \bibnamefont{Ozawa}}, \bibinfo
  {author} {\bibfnamefont{R.}~\bibnamefont{Pantoja}}, \bibinfo {author}
  {\bibfnamefont{E.~A.}\ \bibnamefont{Axtell}},\ and\ \bibinfo {author}
  {\bibfnamefont{S.~M.}\ \bibnamefont{Kauzlarich}},\ }%
  \bibfield{journal}{%
  \bibinfo {journal} {J. Solid State Chem.}\ }%
  \textbf{\bibinfo {volume} {153}},\ \bibinfo {pages} {275} (\bibinfo {year}
  {2000})%
  \bibAnnoteFile{NoStop}{ozawa:00}%
\bibitem{ozawa:04}%
  \BibitemOpen
  \bibfield{author}{%
  \bibinfo {author} {\bibfnamefont{T.~C.}\ \bibnamefont{Ozawa}}\ and\ \bibinfo
  {author} {\bibfnamefont{S.~M.}\ \bibnamefont{Kauzlarich}},\ }%
  \bibfield{journal}{%
  \bibinfo {journal} {J. Cryst. Growth}\ }%
  \textbf{\bibinfo {volume} {265}},\ \bibinfo {pages} {571} (\bibinfo {year}
  {2004})%
  \bibAnnoteFile{NoStop}{ozawa:04}%
\bibitem{axtell:97}%
  \BibitemOpen
  \bibfield{author}{%
  \bibinfo {author} {\bibfnamefont{E.~A.}\ \bibnamefont{Axtell}}, \bibinfo
  {author} {\bibfnamefont{T.}~\bibnamefont{Ozawa}},\ and\ \bibinfo {author}
  {\bibfnamefont{S.~M.}\ \bibnamefont{Kauzlarich}},\ }%
  \bibfield{journal}{%
  \bibinfo {journal} {J. Solid State Chem.}\ }%
  \textbf{\bibinfo {volume} {134}},\ \bibinfo {pages} {423} (\bibinfo {year}
  {1997})%
  \bibAnnoteFile{NoStop}{axtell:97}%
\bibitem{liu:09}%
  \BibitemOpen
  \bibfield{author}{%
  \bibinfo {author} {\bibfnamefont{R.~H.}\ \bibnamefont{Liu}}, \bibinfo
  {author} {\bibfnamefont{D.}~\bibnamefont{Tan}}, \bibinfo {author}
  {\bibfnamefont{Y.~A.}\ \bibnamefont{Song}}, \bibinfo {author}
  {\bibfnamefont{Q.~J.}\ \bibnamefont{Li}}, \bibinfo {author}
  {\bibfnamefont{Y.~J.}\ \bibnamefont{Yan}}, \bibinfo {author}
  {\bibfnamefont{J.~J.}\ \bibnamefont{Ying}}, \bibinfo {author}
  {\bibfnamefont{Y.~L.}\ \bibnamefont{Xie}}, \bibinfo {author}
  {\bibfnamefont{X.~F.}\ \bibnamefont{Wang}},\ and\ \bibinfo {author}
  {\bibfnamefont{X.~H.}\ \bibnamefont{Chen}},\ }%
  \bibfield{journal}{%
  \bibinfo {journal} {Phys. Rev. B}\ }%
  \textbf{\bibinfo {volume} {80}},\ \bibinfo {pages} {144516} (\bibinfo {year}
  {2009})%
  \bibAnnoteFile{NoStop}{liu:09}%
\bibitem{pickett:98}%
  \BibitemOpen
  \bibfield{author}{%
  \bibinfo {author} {\bibfnamefont{W.~E.}\ \bibnamefont{Pickett}},\ }%
  \bibfield{journal}{%
  \bibinfo {journal} {Phys. Rev. B}\ }%
  \textbf{\bibinfo {volume} {58}},\ \bibinfo {pages} {4335} (\bibinfo {year}
  {1998})%
  \bibAnnoteFile{NoStop}{pickett:98}%
\bibitem{huang:13}%
  \BibitemOpen
  \bibfield{author}{%
  \bibinfo {author} {\bibfnamefont{Y.}~\bibnamefont{Huang}}, \bibinfo {author}
  {\bibfnamefont{H.~P.}\ \bibnamefont{Wang}}, \bibinfo {author}
  {\bibfnamefont{W.~D.}\ \bibnamefont{Wang}}, \bibinfo {author}
  {\bibfnamefont{Y.~G.}\ \bibnamefont{Shi}},\ and\ \bibinfo {author}
  {\bibfnamefont{N.~L.}\ \bibnamefont{Wang}},\ }%
  \bibfield{journal}{%
  \bibinfo {journal} {Phys. Rev. B}\ }%
  \textbf{\bibinfo {volume} {87}},\ \bibinfo {pages} {100507(R)} (\bibinfo
  {year} {2013})%
  \bibAnnoteFile{NoStop}{huang:13}%
\bibitem{wang:10}%
  \BibitemOpen
  \bibfield{author}{%
  \bibinfo {author} {\bibfnamefont{X.~F.}\ \bibnamefont{Wang}}, \bibinfo
  {author} {\bibfnamefont{Y.~J.}\ \bibnamefont{Yan}}, \bibinfo {author}
  {\bibfnamefont{J.~J.}\ \bibnamefont{Ying}}, \bibinfo {author}
  {\bibfnamefont{Q.~J.}\ \bibnamefont{Li}}, \bibinfo {author}
  {\bibfnamefont{M.}~\bibnamefont{Zhang}}, \bibinfo {author}
  {\bibfnamefont{N.}~\bibnamefont{Xu}},\ and\ \bibinfo {author}
  {\bibfnamefont{X.~H.}\ \bibnamefont{Chen}},\ }%
  \bibfield{journal}{%
  \bibinfo {journal} {J. Phys.: Condens. Matter}\ }%
  \textbf{\bibinfo {volume} {22}},\ \bibinfo {pages} {075702} (\bibinfo {year}
  {2010})%
  \bibAnnoteFile{NoStop}{wang:10}%
\bibitem{doan:12}%
  \BibitemOpen
  \bibfield{author}{%
  \bibinfo {author} {\bibfnamefont{P.}~\bibnamefont{Doan}}, \bibinfo {author}
  {\bibfnamefont{M.}~\bibnamefont{Gooch}}, \bibinfo {author}
  {\bibfnamefont{Z.}~\bibnamefont{Tang}}, \bibinfo {author}
  {\bibfnamefont{B.}~\bibnamefont{Lorenz}}, \bibinfo {author}
  {\bibfnamefont{A.}~\bibnamefont{M{\"o}ller}}, \bibinfo {author}
  {\bibfnamefont{J.}~\bibnamefont{Tapp}}, \bibinfo {author}
  {\bibfnamefont{P.~C.~W.}\ \bibnamefont{Chu}},\ and\ \bibinfo {author}
  {\bibfnamefont{A.~M.}\ \bibnamefont{Guloy}},\ }%
  \bibfield{journal}{%
  \bibinfo {journal} {J. Am. Chem. Soc.}\ }%
  \textbf{\bibinfo {volume} {134}},\ \bibinfo {pages} {16520} (\bibinfo {year}
  {2012})%
  \bibAnnoteFile{NoStop}{doan:12}%
\bibitem{yajima:12}%
  \BibitemOpen
  \bibfield{author}{%
  \bibinfo {author} {\bibfnamefont{T.}~\bibnamefont{Yajima}}, \bibinfo {author}
  {\bibfnamefont{K.}~\bibnamefont{Nakano}}, \bibinfo {author}
  {\bibfnamefont{F.}~\bibnamefont{Takeiri}}, \bibinfo {author}
  {\bibfnamefont{T.}~\bibnamefont{Ono}}, \bibinfo {author}
  {\bibfnamefont{Y.}~\bibnamefont{Hosokoshi}}, \bibinfo {author}
  {\bibfnamefont{Y.}~\bibnamefont{Matsushita}}, \bibinfo {author}
  {\bibfnamefont{J.}~\bibnamefont{Hester}}, \bibinfo {author}
  {\bibfnamefont{Y.}~\bibnamefont{Kobayashi}},\ and\ \bibinfo {author}
  {\bibfnamefont{H.}~\bibnamefont{Kageyama}},\ }%
  \bibfield{journal}{%
  \bibinfo {journal} {J. Phys. Soc. Jpn.}\ }%
  \textbf{\bibinfo {volume} {81}},\ \bibinfo {pages} {103706} (\bibinfo {year}
  {2012})%
  \bibAnnoteFile{NoStop}{yajima:12}%
\bibitem{singh:12}%
  \BibitemOpen
  \bibfield{author}{%
  \bibinfo {author} {\bibfnamefont{D.~J.}\ \bibnamefont{Singh}},\ }%
  \bibfield{journal}{%
  \bibinfo {journal} {NJP}\ }%
  \textbf{\bibinfo {volume} {14}},\ \bibinfo {pages} {123003} (\bibinfo {year}
  {2012})%
  \bibAnnoteFile{NoStop}{singh:12}%
\bibitem{subedi:13}%
  \BibitemOpen
  \bibfield{author}{%
  \bibinfo {author} {\bibfnamefont{A.}~\bibnamefont{Subedi}},\ }%
  \bibfield{journal}{%
  \bibinfo {journal} {Phys. Rev. B}\ }%
  \textbf{\bibinfo {volume} {87}},\ \bibinfo {pages} {054506} (\bibinfo {year}
  {2013})%
  \bibAnnoteFile{NoStop}{subedi:13}%
\bibitem{bouquet:01}%
  \BibitemOpen
  \bibfield{author}{%
  \bibinfo {author} {\bibfnamefont{F.}~\bibnamefont{Bouquet}}, \bibinfo
  {author} {\bibfnamefont{R.~A.}\ \bibnamefont{Fisher}}, \bibinfo {author}
  {\bibfnamefont{N.~E.}\ \bibnamefont{Phillips}}, \bibinfo {author}
  {\bibfnamefont{D.~G.}\ \bibnamefont{Hinks}},\ and\ \bibinfo {author}
  {\bibfnamefont{J.~D.}\ \bibnamefont{Jorgensen}},\ }%
  \bibfield{journal}{%
  \bibinfo {journal} {Phys. Rev. Lett.}\ }%
  \textbf{\bibinfo {volume} {87}},\ \bibinfo {pages} {047001} (\bibinfo {year}
  {2001})%
  \bibAnnoteFile{NoStop}{bouquet:01}%
\bibitem{lorenz:06}%
  \BibitemOpen
  \bibfield{author}{%
  \bibinfo {author} {\bibfnamefont{B.}~\bibnamefont{Lorenz}}, \bibinfo {author}
  {\bibfnamefont{O.}~\bibnamefont{Perner}}, \bibinfo {author}
  {\bibfnamefont{J.}~\bibnamefont{Eckert}},\ and\ \bibinfo {author}
  {\bibfnamefont{C.~W.}\ \bibnamefont{Chu}},\ }%
  \bibfield{journal}{%
  \bibinfo {journal} {Supercond. Sci. Technol.}\ }%
  \textbf{\bibinfo {volume} {19}},\ \bibinfo {pages} {912} (\bibinfo {year}
  {2006})%
  \bibAnnoteFile{NoStop}{lorenz:06}%
\bibitem{werthamer:66}%
  \BibitemOpen
  \bibfield{author}{%
  \bibinfo {author} {\bibfnamefont{N.~R.}\ \bibnamefont{Werthamer}}, \bibinfo
  {author} {\bibfnamefont{E.}~\bibnamefont{Helfand}},\ and\ \bibinfo {author}
  {\bibfnamefont{P.~C.}\ \bibnamefont{Hohenberg}},\ }%
  \bibfield{journal}{%
  \bibinfo {journal} {Phys. Rev.}\ }%
  \textbf{\bibinfo {volume} {147}},\ \bibinfo {pages} {295} (\bibinfo {year}
  {1966})%
  \bibAnnoteFile{NoStop}{werthamer:66}%
\bibitem{padamsee:73}%
  \BibitemOpen
  \bibfield{author}{%
  \bibinfo {author} {\bibfnamefont{H.}~\bibnamefont{Padamsee}}, \bibinfo
  {author} {\bibfnamefont{J.~E.}\ \bibnamefont{Neighbor}},\ and\ \bibinfo
  {author} {\bibfnamefont{C.~A.}\ \bibnamefont{Shiffman}},\ }%
  \bibfield{journal}{%
  \bibinfo {journal} {J. Low. Temp. Phys.}\ }%
  \textbf{\bibinfo {volume} {12}},\ \bibinfo {pages} {387} (\bibinfo {year}
  {1973})%
  \bibAnnoteFile{NoStop}{padamsee:73}%
\bibitem{kitagawa:13}%
  \BibitemOpen
  \bibfield{author}{%
  \bibinfo {author} {\bibfnamefont{S.}~\bibnamefont{Kitagawa}}, \bibinfo
  {author} {\bibfnamefont{K.}~\bibnamefont{Ishida}}, \bibinfo {author}
  {\bibfnamefont{K.}~\bibnamefont{Nakano}}, \bibinfo {author}
  {\bibfnamefont{T.}~\bibnamefont{Yajima}},\ and\ \bibinfo {author}
  {\bibfnamefont{H.}~\bibnamefont{Kageyama}},\ }%
  \bibfield{journal}{%
  \bibinfo {journal} {Phys. Rev. B}\ }%
  \textbf{\bibinfo {volume} {87}},\ \bibinfo {pages} {060510(R)} (\bibinfo
  {year} {2013})%
  \bibAnnoteFile{NoStop}{kitagawa:13}%
\bibitem{mcmillan:68}%
  \BibitemOpen
  \bibfield{author}{%
  \bibinfo {author} {\bibfnamefont{W.~L.}\ \bibnamefont{McMillan}},\ }%
  \bibfield{journal}{%
  \bibinfo {journal} {Phys. Rev.}\ }%
  \textbf{\bibinfo {volume} {167}},\ \bibinfo {pages} {331} (\bibinfo {year}
  {1968})%
  \bibAnnoteFile{NoStop}{mcmillan:68}%
\bibitem{biani:98}%
  \BibitemOpen
  \bibfield{author}{%
  \bibinfo {author} {\bibfnamefont{F.~F.}\ \bibnamefont{de~Biani}}, \bibinfo
  {author} {\bibfnamefont{P.}~\bibnamefont{Alemany}},\ and\ \bibinfo {author}
  {\bibfnamefont{E.}~\bibnamefont{Canadell}},\ }%
  \bibfield{journal}{%
  \bibinfo {journal} {Inorg. Chem.}\ }%
  \textbf{\bibinfo {volume} {37}},\ \bibinfo {pages} {5807} (\bibinfo {year}
  {1998})%
  \bibAnnoteFile{NoStop}{biani:98}%
\bibitem{yajima:13}%
  \BibitemOpen
  \bibfield{author}{%
  \bibinfo {author} {\bibfnamefont{T.}~\bibnamefont{Yajima}}, \bibinfo {author}
  {\bibfnamefont{K.}~\bibnamefont{Nakano}}, \bibinfo {author}
  {\bibfnamefont{F.}~\bibnamefont{Takeiri}}, \bibinfo {author}
  {\bibfnamefont{Y.}~\bibnamefont{Nozaki}}, \bibinfo {author}
  {\bibfnamefont{Y.}~\bibnamefont{Kobayashi}},\ and\ \bibinfo {author}
  {\bibfnamefont{H.}~\bibnamefont{Kageyama}},\ }%
  \bibfield{journal}{%
  \bibinfo {journal} {J. Phys. Soc. Jpn.}\ }%
  \textbf{\bibinfo {volume} {82}},\ \bibinfo {pages} {033705} (\bibinfo {year}
  {2013})%
  \bibAnnoteFile{NoStop}{yajima:13}%
\bibitem{zhai:13}%
  \BibitemOpen
  \bibfield{author}{%
  \bibinfo {author} {\bibfnamefont{H.-F.}\ \bibnamefont{Zhai}}, \bibinfo
  {author} {\bibfnamefont{W.-H.}\ \bibnamefont{Jiao}}, \bibinfo {author}
  {\bibfnamefont{Y.-L.}\ \bibnamefont{Sun}}, \bibinfo {author}
  {\bibfnamefont{J.-K.}\ \bibnamefont{Bao}}, \bibinfo {author}
  {\bibfnamefont{H.}~\bibnamefont{Jiang}}, \bibinfo {author}
  {\bibfnamefont{X.-J.}\ \bibnamefont{Yang}}, \bibinfo {author}
  {\bibfnamefont{Z.-T.}\ \bibnamefont{Tang}}, \bibinfo {author}
  {\bibfnamefont{Q.}~\bibnamefont{Tao}}, \bibinfo {author}
  {\bibfnamefont{X.-F.}\ \bibnamefont{Xu}}, \bibinfo {author}
  {\bibfnamefont{Y.-K.}\ \bibnamefont{Li}}, \bibinfo {author}
  {\bibfnamefont{C.}~\bibnamefont{Cao}}, \bibinfo {author}
  {\bibfnamefont{J.-H.}\ \bibnamefont{Dai}}, \bibinfo {author}
  {\bibfnamefont{Z.-A.}\ \bibnamefont{Xu}},\ and\ \bibinfo {author}
  {\bibfnamefont{G.-H.}\ \bibnamefont{Cao}},\ }%
  \bibfield{journal}{%
  \bibinfo {journal} {Phys. Rev. B}\ }%
  \textbf{\bibinfo {volume} {87}},\ \bibinfo {pages} {100502(R)} (\bibinfo
  {year} {2013})%
  \bibAnnoteFile{NoStop}{zhai:13}%
\end{thebibliography}

%

\end{document}